# Silicon Waveguides for High-Speed Optical Transmissions and Parametric Conversion around 2 µm

M. Lamy, C. Finot, P. Colman, J. Fatome, G. Millot, G. Roelkens, B. Kuyken and K. Hammani

*Abstract*—We show that single mode Si waveguides efficiently sustain high-speed transmissions at 2 µm. We report the transmission of a 10 Gbit/s signal over 7 cm with a power penalty below 1 dB. Parametric conversion in the continuous wave regime is also demonstrated using the same waveguide structure with a conversion window over 70 nm and an efficiency as high as -25 dB that can be further increased in the pulsed pump regime.

## I. INTRODUCTION

Nowadays, the optical communication traffic continuously increases approaching inexorably a 'capacity crunch' where the conventional C-band around 1.55 µm will not be sufficient anymore; and alternative approaches must be adopted. Recently, the 2-µm spectral region has been suggested as a new possible transmission window [1], which benefits from the emergence of thulium doped fiber amplifiers (TDFA) with broadband and high gain spanning from 1900 nm to 2100 nm. This trend has stimulated complementary studies dealing with dedicated photonic components [1] such as InP-based modulators [2] or arrayed waveguide gratings [3]. Furthermore, high bit-rate data transmissions over distances exceeding a hundred meters have already been successfully demonstrated in low-loss hollow core bandgap photonic fibers designed to exhibit minimal losses around 2000 nm [4]. In this context, on-chip optical transmissions over much shorter distances (typically from a few hundreds of µm up to several cm) also deserve an interest, especially for on-board connections and photonic routing operations.

Therefore, new experimental investigations to study and evaluate the potential of various materials in this new spectral range are needed. Several of such platforms have been the subject of recent studies. For instance, efficient data transmission at 10-Gbit/s have been reported in a 2.5 cm-long Si-Ge waveguide [5] as well as in a 575 µm-long TiO₂ waveguide [6]. However, despite the increasing interest in mid-infrared silicon photonics [7, 8], it is important to note that no high bit-rate transmission has been reported so far at wavelengths around 2 µm for Si waveguides. Here, we demonstrate, for the first time to our knowledge, an error-free transmission of a 10-Gbit/s on-off keying signal at 1.98 µm in a 7-cm long Si waveguide.

We also discuss the possibility to achieve nonlinear processing under continuous wave (CW) operation. Indeed, in order to fulfill the need of transparent optical networks required for wavelength division multiplexing around 2 µm, frequency conversion is a critical operation that should ideally be realized all-optically. Previous examples have shown the efficiency of a parametric process in tellurite or tapered chalcogenide microstructured fibers around 2 µm [9, 10]. Si platforms also offer various nonlinear effects that can be exploited to implement ultrafast integrated optical devices [11, 12] and is therefore ideally suited for this kind of nonlinear processing. Experiments using a Si waveguide operating in the C-band have already shown impressive results [13, 14] while pioneering works around 2 µm have been also reported in 2010 and 2011 with either picosecond pumping [15, 16] or CW pumping [17, 18]. Here, we show that the same architecture as the one involved for C-band operations is also suitable for parametric wavelength conversion in the 2-µm range, and reveal a conversion efficiency of -25 dB for CW operation (-10 dB in the pulsed regime) and a wide tunability over 70 nm.

## II. WAVEGUIDE DESIGN AND FABRICATION

The compact integrated planar waveguide we exploit in this work is rather similar to the design previously reported in [19, 20]. The silicon photonic wires were fabricated in a CMOS pilot line, using 200 mm SOI wafers consisting of a 220 nm silicon waveguide layer resting on a 2 µm buried oxide layer on top of a silicon handle wafer. The wires are about 900 nm wide without top cladding, as shown in Fig. 1. The waveguides are single-mode around 2 µm and only sustain the propagation of the fundamental quasi-TE mode. A group index velocity of 3.87 is obtained by plane wave expansion method (MPB Package [21]). The high optical index contrast also enables a strong confinement of light within the core of the waveguide, resulting in an effective area $A_{\text{eff}}$ (panel (c)) as small as 0.18 µm² and an

This work is financially supported by PARI PHOTCOM Région Bourgogne, by Carnot Arts Institute (PICASSO 2.0 project), by the Institut Universitaire de France, by FEDER-FSE Bourgogne 2014/2020 and by the french "Investissements d'Avenir" program, project ISITE-BFC (contract ANR-15-IDEX-0003). The research work has benefited from the PICASSO experimental platform of the University of Burgundy.

M. L., C. F. P.C, J. F., G. M. and KH are with the Laboratoire Interdisciplinaire Carnot de Bourgogne, UMR 6303 CNRS-Université Bourgogne-Franche-Comté, 9 av. A. Savary, 21078 Dijon cedex, France (kamal.hammani@u-bourgogne.fr).

B. K. and G.R are with the Photonics Research Group, Ghent University, imec, Technologiepark 15, 9052 Ghent, Belgium (bart.kuyken@ugent.be)

estimated nonlinearity parameter $\gamma = 2\pi\,n_2/\lambda\,A_{\text{eff}}$ of 130/W/m at $\lambda = 1.98$ µm (assuming a nonlinear index $n_2 = 0.57$ m²/W) [22].

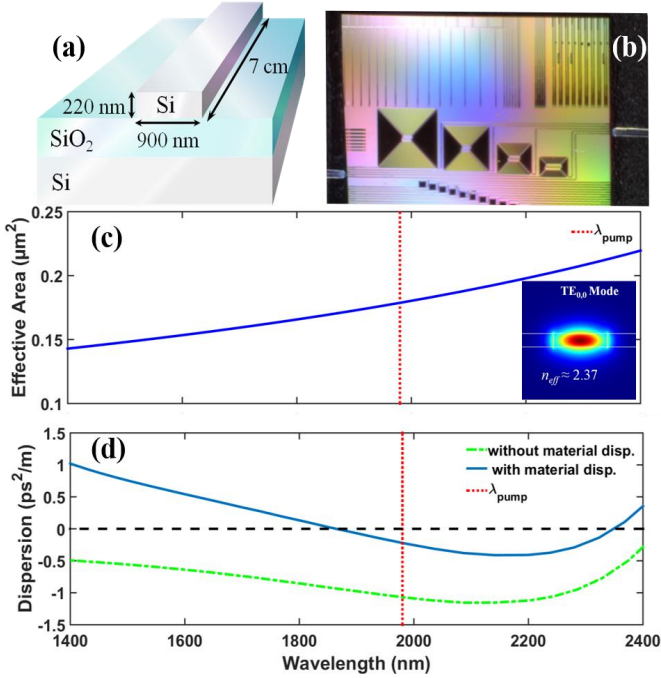

Fig. 1. (a) Structure of the waveguide under investigation. (b) Picture of the optical chip. (c) Effective area of the waveguide. Inset: Detail of the TE mode at the pump wavelength (d) Dispersive properties of the waveguide.

Particular care has been devoted to the optimization of the dispersion profile of the Si waveguide. Details of the spectral evolution of the second order dispersion coefficient are provided in panel (d). In order to reach the highest performance, operation in the anomalous dispersion regime has been favored. Here, the dispersion coefficient is estimated to be around - 0.23 ps²/m at 1980 nm. 2-cm and 7-cm long samples have been investigated. Propagation losses at 2 µm have been estimated around 3.4 dB/cm. Light coupling is achieved through butt-coupling assisted with fibered lenses, hence without any mode adapters. Therefore, typical coupling efficiency is found to be around 10 dB/facet.

## III. DATA TRANSMISSION AT 2 µM

In order to demonstrate the suitability of the Si device for 2-µm optical communications, we have implemented the experimental setup depicted in Fig. 2(a) and based on 2-µm commercially available devices. The transmitter (TX) was based on an intensity modulated laser diode centered at 1980 nm by means of a Niobate-Lithium modulator (IM). The Non-Return-to-Zero On-Off-Keying signal under test was a $2^{31}$-1 pseudorandom bit sequence (PRBS) at 10 Gbit/s. Since the single-mode waveguide was polarization-sensitive, a polarization controller (PC) was inserted after the intensity modulator. Then a first thulium doped fiber amplifier (TDFA) was used to boost the signal before injection into the waveguide. The receiver (RX) was based on a second TDFA. This TDFA was set to work with a constant gain instead of a constant output power as usually used in the C-band. Therefore, a variable optical attenuator was implemented at the TDFA output ensuring that the photodiode was operating at a constant power level. An optical bandpass filter (OBPF, 0.64 nm bandwidth) is also inserted at the output of the system in order to limit the accumulation of amplified spontaneous emission from the TDFAs. The optical signal was finally detected with a photodiode (PD) with an electrical bandwidth of 12.5 GHz. An optical spectrum analyzer (OSA) was used to evaluate the optical signal to noise ratio (OSNR).

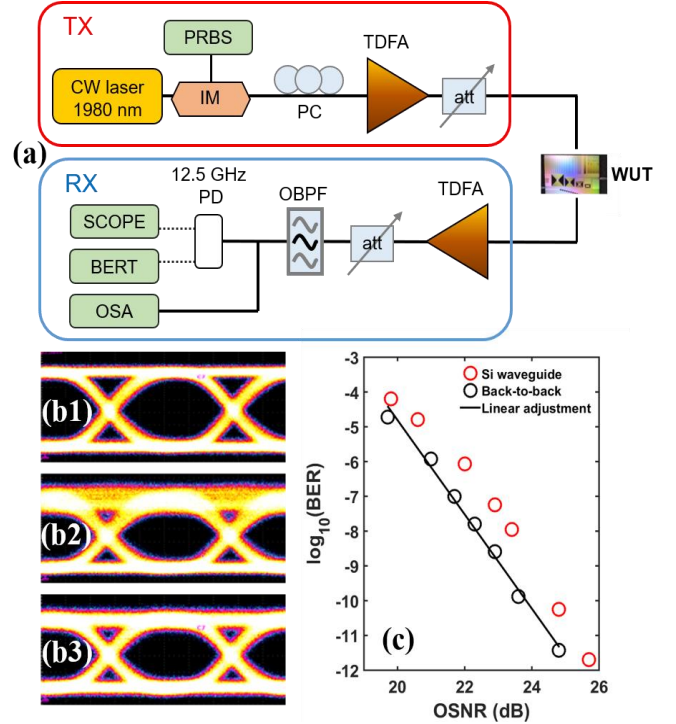

Fig. 2. (a) Experimental setup (b) 10-Gbit/s eye diagrams recorded in the back-to-back configuration, after propagation in a 2-cm and 7-cm long waveguide (panel 1, 2 and 3 respectively). (c) Bit-error-rate measurements (BER) as a function of OSNR for the 7-cm long Si waveguide.

Figs. 2(b) illustrate the results of the 10-Gbit/s transmission at 2 µm by providing examples of the corresponding eye-diagrams for the back-to-back measurement (b1) as well as at the output of the 2-cm and 7-cm long waveguides (b2 and b3, respectively). In both transmission experiments, widely open eyes were obtained, meaning that the insertion of the waveguides under test did not induce any major impairments on the transmission. However, when comparing the level of '1' symbols obtained with the 2-cm waveguide (b2) with the back-to-back results (b1), we observed a rather thick level that was not present for the 7-cm waveguide. This amplitude jitter is definitively not linked to an excessive noise level, but is ascribed to a Fabry-Perot effect occurring between the waveguide end-facets. Indeed, the low level of attenuation into the waveguide combined with the high Fresnel reflection coefficient caused by the large index contrast between Si and air results into interference pattern between the signal and its replica delayed by 520 ps, thus affecting the level of the '1' bit. Note that simple treatments of the input and output facets of the waveguide, e.g. depositing thin layers as an anti-reflective coating could efficiently reduce this interference process. We then focus our work on longer propagation distances and can observe in Fig. 2(b3) that for the 7-cm long waveguide, increasing losses tends to mitigate this interference impairment, thus leading to fluctuations of the '1' level similar to those of the '0' level. Error-free transmission at 2 μm can then be successfully achieved through this Si waveguide. The quality of the 10-Gbit/s transmitted signal was more quantitatively assessed through systematic measurements of the bit-error-rate according to the OSNR on the receiver. Results are summarized in Fig. 2(c). A moderate penalty (~1 dB) was obtained after transmission through the waveguide and compared to the back-to-back configuration.

## IV. WAVELENGTH CONVERSION AT 2 μM

In this section, we aim to evaluate the potential of our Si waveguide for wavelength conversion applications. Details of the experiment are provided in Fig. 3(b). A CW laser tunable between 1925 and 2010 nm was used to generate a signal wave, while the pump beam is kept constant to 1980 nm. By means of an intensity modulator driven by an electrical pulse pattern generator, we were able to study the wavelength conversion process in either continuous or pulsed regime with a typical pump duration of 100 ps. The delay between two consecutive pulses can be adjusted from 200 ps to 1.6 ns, corresponding to repetition rates between 625 MHz and 5 GHz.

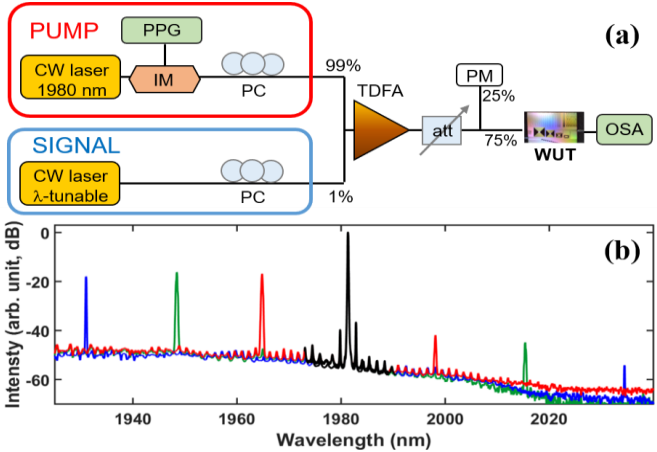

Fig. 3. (a) Experimental setup involved in the wavelength conversion process. (b) Wavelength conversion achieved in the CW regime for a signal detuned by 16, 32 and 50 nm (red, green and blue curves respectively). The black line represents the pump spectrum.

The pump and signal waves were combined using a 99/1 coupler before being simultaneously amplified by a TDFA. The optical power at the waveguide input was adjusted with a variable attenuator and monitored by a power meter. We focus here more particularly on the 2-cm long waveguide which is already twice as long as the effective length. At the output of the system, an OSA was used to evaluate the conversion efficiency defined as the ratio between the output powers of the signal and idler waves.

Three examples of wavelength conversion achieved for different wavelength detunings are shown in Fig. 3(b) for an estimated input CW pump power of 12 dBm. The conversion efficiency of the four-wave mixing process can be as high as -25 dB and can be tuned over a bandwidth of several tens of nm. A more exhaustive study is reported in Fig. 4(a) showing an efficient conversion process of the idler wave detuned up to 60 nm from the pump. Note that below 1960 nm, our experiment was only limited by the tunability range of the signal wavelength. We also notice a drop in the conversion efficiency around 2028 nm which is typically related to the phase matching condition of the parametric process under-study. Furthermore, the large bandwidth achieved here denotes that, as expected, the pump lies in the vicinity of the zero-dispersion wavelength. Finally, the conversion efficiency was found slightly higher for low-wavelength idler waves. Such an asymmetry can be attributed to the impact of the ASE noise of the TDFAs. In order to assess the reproducibility of the performances, we also tested a second 2-cm long waveguide on *spare* chip with the similar physical properties. Results in red on Fig. 4(a) show the fair reproducibility of these results.

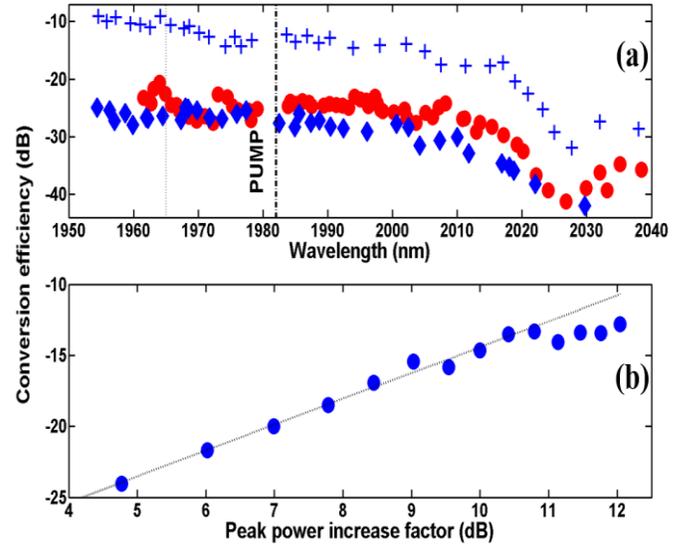

Fig. 4. (a) Evolution of the conversion efficiency as a function of the idler wavelength. Results were recorded for 2-cm long samples. Results achieved in CW pumping regime are plotted with circles and diamonds whereas the pulsed regime is reported with crosses (duty-cycle of 1/16). (b) Evolution of the conversion efficiency versus peak power of the pulsed signal for a converted wavelength of 1965 nm (blue circles), dashed-line depicts for a logarithmic fit.

Finally, we have also tested the conversion efficiency when using a pulsed signal. This allowed, for a given average power, to reach a higher peak power $P_P$. Results recorded for a signal wavelength of 1995 nm are summarized on panel 4(b) and show that conversion efficiencies close to -10 dB can be achieved for a peak-power around 250 mW. Performing a logarithmic fit of these results (dashed-line in 4b), shows that the conversion efficiency scales as $P_P^{1.82}$ which is in qualitative agreement with the exponent 2 expected for usual four-wave mixing process in absence of higher-order nonlinear effects. Note that although the detrimental two photon absorption around 2.0 μm was strongly reduced compared to 1550 nm, this nonlinear absorption does not completely vanish [23].

## V. CONCLUSION

To conclude, we demonstrated that single-mode Si waveguides can sustain error-free transmission of high-speed telecom signals around 2 µm on distance up to 7 cm with a very moderate penalty. We have also shown that the same waveguide design, optimized to reach low level of chromatic dispersion around 2 µm, is suitable for wavelength conversion process with a wavelength bandwidth as high as 70 nm and an conversion efficiency up to -10 dB when operating in pulsed regime. Furthermore, given the moderate level of maturity of the various 2-µm components that are to date available, such characterization of telecommunication signals are not as straightforward as in the C-band. Consequently, it was not possible in the present study to perform a wavelength conversion process on genuine encoded data and to evaluate the quality of such a converted signal. However, both the error-free transmission experiments as well as the wavelength conversion study confirm that the 2-µm spectral window is a promising alternative deserving further investigations in a near future, especially in the context of high speed optical processing based on low-cost and ultra-compact optical chips combined with electronic circuits [24].